# Einstein contra la mecánica cuántica
*…el azar, la ignorancia y nuestra ignorancia sobre el azar…*[*]

Juan Pablo Paz

*Departamento de Física "Juan José Giambiagi", FCEyN, UBA,
Pabellón 1 Ciudad Universitaria, 1428 Buenos Aires, Argentina.*

Einstein nunca pudo reconciliarse con la mecánica cuántica que es a la vez la teoría física más exitosa y más anti-intuitiva de la historia. La crítica más aguda contra esta teoría fue presentada por Einstein en 1935 en un célebre trabajo publicado junto con sus colaboradores Podolsky y Rosen (EPR). El cuestionamiento de EPR no dio lugar al derrumbe de la mecánica cuántica sino que permitió exhibir con toda crudeza las extrañas propiedades de esta teoría. Los avances de la física de fines del siglo XX demostraron que las ideas de EPR sobre lo incompleta que era la mecánica cuántica eran incorrectas. En este trabajo resumimos los ingredientes principales de la mecánica cuántica y exponemos los cuestionamientos de Einstein hacia ella. Nos concentramos en el análisis de la paradoja de EPR y en la forma en la que puede verificarse la validez de la mecánica cuántica a través de la detección de violaciones a las desigualdades de Bell. Finalmente, resumimos algunos avances recientes que apuntan a usar las propiedades más extrañas de la física cuántica para desarrollar tecnologías que podrían modificar la transmisión y el procesamiento de la información en el siglo XXI.

---

## I. Introducción

Albert Einstein fue, sin duda, uno de los más grandes científicos de la historia. Sus ideas revolucionaron el pensamiento humano mostrando que, por ejemplo, conceptos tan básicos como el tamaño de los objetos y la duración de los intervalos de tiempo no tienen un carácter absoluto sino que, por el contrario, dependen del observador. La Teoría de la Relatividad nos obligó a repensar conceptos básicos que están anclados en nuestro sentido común, ese conjunto de ideas que consideramos como obvias y que se forjan en nuestras experiencias cotidianas. Hoy, a más de cien años del nacimiento de la Relatividad hay miles de científicos que la comprenden y aplican a la perfección. Más aún, a lo largo de estos cien años ha sido posible reconciliar algunas de sus ideas más extrañas con nuestro sentido común. Para ello muchas veces solemos apelar a metáforas, como por ejemplo: la vida en un espacio curvo puede imaginarse pensando en lo que le ocurriría a un ser plano condenado a existir sobre la superficie de una esfera; o bien la materia curva el espacio-tiempo de manera análoga a como una cama elástica se deforma al apoyar sobre ella un objeto masivo, etc. Estas analogías tienen sus defectos y no siempre resultan precisas, pero al menos es posible imaginarlas.[1]

La teoría de la relatividad se desarrolló en sus comienzos como una empresa familiar: Einstein la concibió trabajando en soledad. Es bien conocida la anécdota que cuenta que cerca de 1920 Sir Arthur Eddington fue reporteado

---

[1] La influencia de Einstein en el pensamiento científico moderno se pone en evidencia mencionando algunas frases llamativas que no tendrían sentido sin su contribución: "el tiempo se dilata", "las longitudes se contraen", "vivimos en un espacio-tiempo curvo", "la luz modifica su trayectoria al pasar cerca del sol", etc

por un periodista que le mencionó el rumor que por ese entonces afirmaba que en el mundo solamente había tres personas que comprendían la teoría de la relatividad. Eddington (¿bromeando?) preguntó: "¿Y quién es el tercero?" Por esos años había otra rama de la física en pleno desarrollo: la física cuántica. Contrariamente a lo que sucedía con la relatividad, eran decenas los físicos que trabajaban activamente en su desarrollo. La mecánica cuántica fue una creación colectiva que surgió luego de un esfuerzo material e intelectual impresionante. En este artículo nos referiremos a ese capítulo de la física, la física cuántica. En particular nos referiremos a los cuestionamientos de Albert Einstein hacia la mecánica cuántica.

Einstein, con su trabajo sobre el efecto fotoeléctrico, contribuyó sustancialmente al desarrollo de la mecánica cuántica. Sin embargo, jamás fue capaz de aceptar sus consecuencias y siempre la aborreció. Para citar solamente alguna de sus opiniones escritas basta mencionar las siguientes. En 1912 en una carta a Heinrich Zangger afirmaba, de manera algo irreverente: "Cuanto más éxitos logra, más tonta me parece". Más adelante, en 1930 en una carta dirigida a Max Born decía: "Todavía no me resigno a creer que los métodos estadísticos de la mecánica cuántica sean la última palabra, pero por el momento soy el único que sostiene esa opinión". En otra carta dirigida a Max Born, Einstein acuñó, en 1944, su famosa frase: "Usted cree que Dios juega a los dados, mientras que yo creo en la existencia de leyes y de orden en un mundo al que, de una manera brutalmente especulativa, estoy tratando de comprender". En 1950, hacia el final de su vida y en una época de gloria de la física cuántica, Einstein se atrevió a afirmar que "...a pesar de sus notables avances parciales, el problema está lejos de tener una solución satisfactoria".

¿Qué era lo que más le molestaba a Einsten de la física cuántica? La respuesta es sencilla: Su insatisfacción se originaba en el indeterminismo. La mecánica cuántica es una teoría no-determinista. Afirma que es posible realizar muchas veces el mismo experimento y obtener siempre resultados diferentes. Para colmo de males, la mecánica cuántica afirma que el indeterminismo es de naturaleza fundamental y que no se origina en ninguna limitación de nuestro instrumental. Es decir, de acuerdo a ella, la razón por la cual al repetir un experimento obtenemos resultados diferentes no es la falta de precisión en los artefactos que utilizamos para preparar el objeto antes de efectuar la medición, ni tampoco la falta de control en los aparatos de medición. Por último, y esto resultaba intolerable para Einstein, la mecánica cuántica afirma que el indeterminismo tampoco puede atribuirse a nuestra ignorancia sobre los detalles del objeto estudiado. Einstein hubiera aceptado de buena gana una teoría que, con modestia, se limitara a predecir probabilidades. En cambio, le resultaba intolerable la mecánica cuántica que de manera contundente, afirma que las probabilidades no surgen de nuestra ignorancia ni de nuestra incapacidad de controlar todas las variables experimentales sino que tienen un origen fundamental e inexplicable.

Estas características de la física cuántica no solamente le molestaban a Einstein, sino que todavía molestan a casi todos los físicos cuánticos, que se cuentan por decenas de miles. Paradójicamente, siendo la física cuántica la teoría científica mejor testeada de la historia, todavía no se han acallado los debates sobre su interpretación. Estos debates comenzaron desde la concepción de la teoría y Einstein tuvo un notable protagonismo en muchos de ellos. Las predicciones de la física cuántica son múltiples y sumamente precisas. Por ejemplo, puede predecir que cada vez que iluminemos un átomo de Helio se emitirá un electrón siempre que la longitud de onda de la luz sea menor que $50.425931 \pm 0.000002$ nanómetros. Por otra parte, esta predicción teórica es contrastada con el resultado de los experimentos donde se comprueba que los electrones son emitidos para longitudes de onda menores que $50.4259299 \pm 0.0000004$ nanómetros. El acuerdo entre la teoría y el experimento es notable: ¡una precisión comparable a la que tendríamos si fuéramos capaces de predecir la distancia entre Ushuaia y La Quiaca con un error menor que diez centímetros![2] Predecir propiedades de los

---

[2] Por el contrario, la física "clásica" predice que para cualquier longitud de onda algunos electrones serán

átomos con precisión asombrosa puede ser impresionante pero alejado de la vida cotidiana. Sin embargo, a partir de este tipo de logros es que la física cuántica ha permitido el desarrollo de tecnologías que cambiaron el mundo y nuestra forma de vida: sin ella no se hubiera desarrollado la energía nuclear, ni la microelectrónica, ni el láser, ni ninguna de las tecnologías optoelectrónicas que revolucionaron las comunicaciones, ni las técnicas modernas de diagnóstico médico por imágenes, etc. ¡Casi todas las tecnologías relevantes del siglo XX se basan en la mecánica cuántica!

Sin embargo, pese a sus asombrosas predicciones, ninguno de las decenas de miles de científicos cuánticos es capaz de "comprender" esta teoría. No es capaz de tornarla compatible con el sentido común.
Richard Feynman, uno de los científicos más brillantes de la segunda mitad del siglo XX afirmaba, en forma provocadora, que "nadie entiende la mecánica cuántica". Y lo hacía en el contexto de una reflexión profunda: para Feynman, nadie es capaz de hacerse una imagen correcta del mundo microscópico usando los conceptos que generamos para describir el mundo macroscópico. Al hacer eso, caemos inevitablemente en preguntarnos, ¿cómo es posible que la naturaleza se comporte de este modo? Nadie lo entiende. Pero los hechos confirman que la naturaleza se comporta tal como lo predice la mecánica cuántica.

Este trabajo está organizado de la siguiente manera. En la Sección 2 presentaremos los ingredientes básicos de la física cuántica. Intentaremos hacer una presentación desprovista de tecnicismos matemáticos. Comenzaremos por describir los famosos principios de complementariedad e incertidumbre. Luego discutiremos brevemente las curiosas predicciones cuánticas para los sistemas compuestos, para lo que introduciremos la noción de entrelazamiento. En la Sección 3 presentaremos la crítica crucial que Einstein realizó junto con sus colaboradores Boris Podolsky y Nathan Rosen en un celebrado trabajo publicado en 1935 (y que, paradójicamente, ¡se ha convertido en el trabajo científico mas citado de todos los publicados por Einstein!).

En la Sección 4 presentaremos un argumento formulado originalmente por John Bell, y que en su momento fue considerado como uno de los trabajos más importantes de la historia de la ciencia: mostraremos que es posible discernir experimentalmente entre las predicciones cuantitativas que surgen de todas las teorías en las que el azar se origina en la ignorancia y las predicciones realizadas por la mecánica cuántica. En la Sección 5, resumiremos los resultados de los experimentos que confirman las predicciones cuánticas y que invalidan a un enorme conjunto de modelos alternativos que comúnmente se denominan "teorías realistas-locales". En la Sección 6, resumiré los esfuerzos recientes para utilizar las propiedades más anti intuitivas de la mecánica cuántica para desarrollar nuevas tecnologías que, en el futuro, pueden afectar drásticamente la forma en que almacenamos, transmitimos y procesamos la información.

## II. Mecánica Cuántica

La mecánica cuántica nació hace más de un siglo. Lo hizo de manera turbulenta cuando un grupo cada vez más grande de físicos tomó conciencia de que la emisión y absorción de la luz por la materia no podía ser comprendida dentro del marco de las leyes de la física formuladas hasta ese momento. Por esa época reinaban sobre la física el electromagnetismo de Maxwell, la mecánica de Newton y la termodinámica de Boltzmann. La formulación de la nueva mecánica fue una tarea titánica que recayó en personalidades como Planck, Einstein, Bohr, Heisenberg, Schrödinger, Dirac, Fermi, de Broglie, von Neumann, Born, Pauli y muchos otros. El desarrollo de esta teoría comenzó en 1900 y concluyó cerca de 1930, cuando adquirió finalmente coherencia y solidez internas. Sin embargo, los debates sobre los fundamentos y la interpretación de la mecánica cuántica no se han acallado y muchos todavía consideran que existen problemas abiertos, como el famoso "problema de la medición".

Teniendo presente que la mecánica cuántica ya cuenta con su mayoría de edad evitaremos

---

emitidos por los átomos de Helio, lo cual entra en abierta contradicción con los resultados de los experimentos utilizar aquí un enfoque histórico para presentarla. Tal enfoque puede ser encontrado en la mayoría de los libros de divulgación científica (o en la mayor parte de los libros de texto) escritos hasta el presente. Por el contrario, apelaremos a una introducción "brutal" describiendo las bases conceptuales y los aspectos mas anti-intuitivos de la física cuántica. El lector deberá creer que esta construcción teórica no es usada por los físicos debido a su naturaleza perversa. La razón para el éxito de la mecánica cuántica es mucho más pragmática: es el único marco teórico capaz de predecir los resultados de los experimentos que se realizan cotidianamente en nuestros laboratorios.

Algunos autores afirman que en la actualidad vivimos una segunda revolución cuántica. Después de más de un siglo de física cuántica recién ahora existen las tecnologías como para ponerla a prueba hasta sus últimas consecuencias y para aprovecharla para desarrollar una nueva generación de tecnologías. ¿Qué es lo que ha cambiado en estos años? Si bien la física cuántica surgió como un modelo para explicar el comportamiento de los átomos, su desarrollo se basó en experimentos, como los de emisión y absorción de luz por la materia, que siempre involucraban cantidades macroscópicas de átomos.[3] Durante mucho tiempo la manipulación de átomos o moléculas de a una a la vez pareció algo totalmente inconcebible. Inclusive algunos renombrados autores pensaron que esta era una limitación de principio. Argumentaban que, por su carácter probabilístico, la mecánica cuántica solamente era aplicable a conjuntos de muchos sistemas idénticos. Pero en la actualidad, la mecánica cuántica está siendo puesta a prueba (y confirmada en sus aspectos mas anti-intuitivos) en experiencias que involucran cantidades cada vez mas pequeñas de átomos que son manipulados individualmente. Por ejemplo, en las actuales "trampas de iones" es posible manipular átomos de a uno, ubicarlos en lugares predeterminados del espacio y someterlos a todo tipo de análisis. Las más notables predicciones cuánticas que involucran fenómenos como el entrelazamiento no-local entre las partes de un sistema compuesto no solamente han sido comprobadas sino que han abierto la puerta a nuevas tecnologías que seguramente modificarán la forma en la que concebimos el procesamiento y la transmisión de la información. La segunda revolución cuántica es la que viene de la mano de la preparación y el aprovechamiento del entrelazamiento a escala macroscópica. Tal vez esa revolución nos permita avanzar en la incorporación de algunas de las ideas más raras de la mecánica cuántica a nuestro sentido común.

### A. Complementariedad e incertidumbre

Quien conozca la obra teatral "Copenhague", escrita por Michael Frayn, recordará las intensas discusiones entre Niels Bohr y Werner Heisenberg, que en Buenos Aires fueron interpretados magistralmente por Juan Carlos Gene y Alberto Segado. Bohr y Heisenberg discutían sobre la complementariedad y la incertidumbre. Estos son dos de los ingredientes básicos de la mecánica cuántica, que ponen de manifiesto cuán extraño es el comportamiento de la naturaleza a escala microscópica. El *principio de complementariedad* es un verdadero atentado contra nuestra intuición. En su versión más general afirma lo siguiente: *si preparamos un objeto de manera tal que la propiedad A toma un valor preciso, entonces siempre existe otra propiedad B cuyo valor está completamente indeterminado. En ese caso, afirmamos que las propiedades A y B son "complementarias"*. Tal vez la formulación que hemos elegido presentar aquí suena un poco abstracta pero vale la pena pensar un poco sobre su contenido.

El principio se aplica a situaciones muy habituales en las que sometemos a un objeto a algún proceso de preparación tal que si posteriormente medimos repetidamente la propiedad A siempre obtenemos el mismo valor. Lo sorprendente es que el principio de complementariedad afirma que "entonces, siempre existe otra propiedad B cuyo valor esta completamente indeterminado". ¿Qué quiere decir esto? Simplemente significa que si preparamos el sistema en un estado en el que la propiedad A tiene un valor preciso y medimos la propiedad B entonces obtendremos resultados completamente aleatorios. Si repetimos muchas veces este procedimiento (es decir, preparamos el sistema con un valor de A y medimos la propiedad B) obtendremos resultados diferentes, distribuidos de manera totalmente azarosa.

---

[3] ¡En un centímetro cúbico de aire hay cerca de un billón de billones de átomos!

Construir una teoría sobre la base de un principio como este parece un verdadero acto de renunciamiento intelectual. En efecto, nunca antes de la mecánica cuántica la física se había planteado una limitación epistemológica de este tipo. Siempre se había pensado que los objetos que componen el Universo no solamente pueden ser caracterizados por propiedades mensurables (o sea, propiedades que toman valores susceptibles de ser medidos experimentalmente). También, la física siempre aceptó aquello cuya validez resulta obvia a partir de nuestro sentido común: todas las propiedades de un objeto deberían poder determinarse simultáneamente. Por supuesto, la determinación simultánea de los valores de todas las propiedades de un objeto podría ser una tarea técnicamente difícil. Pero las dificultades técnicas o instrumentales son siempre vistas como desafíos, como obstáculos que se pueden superar. En cambio, el principio de complementariedad nos habla de otra cosa: nos enfrenta a una limitación de principio. Nos dice que no todas las propiedades de un objeto son compatibles entre sí. Los valores de las propiedades complementarias no pueden ser determinados simultáneamente si no que son como dos caras de un objeto que nunca pueden ser vistas al mismo tiempo.

El proceso que llevó a que los físicos se vieran forzados a aceptar la validez de un principio tan desagradable como el de complementariedad fue largo y plagado de debates. Pero fue el único remedio para poder formular una teoría cuyas predicciones estuvieran de acuerdo con los resultados de los experimentos. El ejemplo mas conocido de propiedades complementarias es el de la posición de un objeto, a la que denotaremos con la letra "*R*", y su momento, denotado con la letra "*P*" (el momento de un objeto es el producto de su masa por su velocidad). Posición y momento son variables complementarias. Esto contradice todo lo establecido por la física clásica, la física de Newton, que establece que un objeto siempre puede caracterizarse por su posición y su velocidad. La velocidad nos dice como se modifica la posición con el tiempo. De esa forma, al moverse todo objeto describe una trayectoria. En cambio, la mecánica cuántica nos dice que para comprender el mundo microscópico debemos abandonar la idea de que las partículas evolucionan siguiendo trayectorias. Si no renunciamos a las trayectorias no podremos explicar ninguno de los famosos experimentos donde se observa la interferencia de ondas de materia. Es necesario aceptar que en el mundo microscópico cuando las partículas se dirigen desde una fuente hasta un detector no siguen trayectorias bien definidas sino que se deslocalizan, se desdoblan y siguen todas las trayectorias posibles. Este es un fenómeno raro y anti-intuitivo, pero no hay más remedio que aceptarlo para poder comprender la curiosa naturaleza del mundo microscópico.

El *principio de incertidumbre* está íntimamente relacionado con el de complementariedad. En algún sentido es la versión cuantitativa del anterior. Se aplica a situaciones en las que preparamos un objeto en un estado en el que su posición *R* no toma un valor preciso sino que cuando lo medimos obtenemos valores distribuidos con una dispersión *ΔR* alrededor del valor más probable.[4] En una situación de este tipo, si medimos el momento del objeto tampoco obtendremos siempre el mismo valor sino que comprobaremos que los resultados tienen una dispersión *ΔP*. El principio de incertidumbre establece que: *la posición R y el momento P de un objeto son propiedades complementarias y sus varianzas ΔR y ΔP satisfacen la siguiente relación*:

$$\Delta R \Delta P \geq \hbar / 2$$

En esta desigualdad, $\hbar$ está relacionada con la famosa constante de Planck y tiene el valor: $\hbar = 1.05 \times 10^{-34}$ kg m$^2$/s. La desigualdad matemática tiene un impacto profundo: como el

---

[4] Los estadísticos caracterizan una situación como esta diciendo que ΔR es la "varianza" de la distribución de los resultados de la medición de *R*

producto de las dos dispersiones debe ser mayor que una cierta cantidad entonces debe cumplirse que cuanto más pequeña sea la varianza en la posición $\Delta R$, más grande debe ser el valor de la varianza en momento $\Delta P$ (y viceversa). La pequeñez del valor de $\hbar$ (un número con treinta y tres ceros detrás del punto decimal) explica el motivo por el cual las consecuencias de los principios de complementariedad e incertidumbre no son perceptibles en la escala macroscópica. Por ejemplo, si preparamos una partícula de 1 gramo en un estado donde la posición está determinada con una incerteza de $\Delta R$ =1cm, entonces el principio de incertidumbre establece que nunca podremos determinar la velocidad con una incerteza menor que $10^{-28}$ m/seg. Claramente ningún instrumento de medición es capaz de detectar una desviación tan pequeña.

No es posible dejar de sorprenderse por las implicancias de los principios de complementariedad y de incertidumbre, que fueron establecidos respectivamente por Niels Bohr y Werner Heisenberg alrededor de 1925. Ponen en evidencia cuan extraña es la mecánica cuántica y es imposible aceptarlos sin antes intentar demolerlos: Einstein, y cualquier persona en su sano juicio, preguntaría: "¿Cómo es posible que podamos preparar un objeto de modo tal que la propiedad $A$ tiene un valor preciso, pero que sea imposible darle un valor preciso a otra propiedad complementaria $B$? Esta pregunta NO tiene respuesta dentro de la mecánica cuántica. Dicha teoría acepta este hecho sorprendente como una propiedad de la naturaleza y a partir de eso formula un modelo que tiene una notable capacidad predictiva. Con todo dramatismo, la mecánica cuántica se yergue hoy, a más de cien años de su nacimiento, como la única teoría compatible con los resultados experimentales modernos.

### B. ¿Indeterminismo o ignorancia?

A lo largo del siglo XX los físicos hicieron numerosos intentos por encontrar alternativas a la mecánica cuántica y desarrollar teorías que fueran más aceptables para nuestro sentido común. La clase de modelos que naturalmente podrían competir con la mecánica cuántica incluye a aquellos en los que la complementariedad no es una propiedad fundamental sino que es fruto de nuestras limitaciones. Por ejemplo, podríamos imaginar que la naturaleza es tal que cada vez que fijamos el valor de alguna propiedad $A$ perturbamos el objeto de manera tal que afectamos el valor de $B$. En un mundo como ese, la razón por la cual una medición de $B$ da lugar a resultados aleatorios es nuestra incapacidad de controlar todas las propiedades de los objetos o, equivalentemente, nuestra ignorancia sobre detalles del mundo microscópico que todavía son inaccesibles a nuestras limitadas posibilidades experimentales. Einstein, y cualquier persona razonable, hubieran estado dispuestos a aceptar un mundo de estas características. En ese caso, la mecánica cuántica no proveería una descripción completa de la naturaleza sino solamente daría una descripción parcial. En la próxima Sección, presentaremos un famoso argumento formulado por Einstein en 1935 que intentaba demostrar precisamente esto: que la descripción del mundo provista por la mecánica cuántica es incompleta. Más adelante veremos cómo, sorprendentemente, los notables avances de la física de fines del siglo XX fueron capaces de demostrar la falsedad del argumento de Einstein. Es notable, pero la física ha sido capaz de demostrar que el azar no se origina en nuestra ignorancia. Sin embargo, hasta ahora debemos reconocer nuestra ignorancia sobre las causas que originan el azar.

Por último, es oportuno aclarar que los principios de complementariedad e incertidumbre discutidos más arriba no forman parte de los postulados básicos en los que se funda la versión moderna de la mecánica cuántica. En efecto, estos no son verdaderos "principios" sino consecuencias de axiomas todavía más fundamentales que debido a su complejidad matemática no serán discutidos aquí.

### C. El spin, la más cuántica de las propiedades

El spin es una propiedad de algunas partículas que fue descubierta en 1922 en experimentos realizados por Otto Stern y Wolfgang Gerlach.

Una partícula con spin lleva consigo un pequeño imán que, como todo imán, tiene dos polos y puede describirse utilizando una flecha imaginaria (un vector) que se dirige desde el polo sur hacia el norte del mismo. La longitud de la flecha (el módulo del vector) es proporcional a la intensidad del imán. Como veremos, imaginar al spin como una flecha es una sobre-simplificación ya que este personaje tiene muchas características sorprendentes. La más sorprendente de todas ellas es la siguiente: cuando medimos la proyección del spin a lo largo de una dirección cualquiera obtenemos sólo dos valores posibles. Conviene detenerse un poco a pensar en este resultado. La proyección del spin en la dirección $\hat{z}$ es la longitud de la sombra proyectada por la flecha que representa al spin a lo largo de esa dirección. Lo que acabamos de afirmar equivale a decir que siempre que medimos el tamaño de la sombra del spin en cualquier dirección obtenemos solamente dos valores. Además, resulta que estos dos valores son de igual magnitud y de signo contrario (es conveniente notar que la proyección de un vector sobre un eje tiene signo positivo o negativo según cual sea la dirección hacia la que se dirija la proyección de la punta de la flecha).[5] Como dijimos, el tamaño de la flecha que representa al spin está relacionado con la intensidad del imán. Por lo tanto, su valor no se mide en metros sino en otras unidades que resultan ser las mismas que aparecen en la famosa constante de Planck $\hbar$. En efecto, los dos valores que se obtienen a partir de la medición de cualquier componente del spin de una partícula como el electrón (o cualquier otra de las llamadas "partículas de spin 1/2") son siempre $+\hbar/2$ y $-\hbar/2$.

Para medir la proyección del spin a lo largo de una dirección cualquiera necesitamos un aparato como el descripto en la Figura 1. Dicho aparato mide la proyección del spin sobre el eje que en la Figura denominamos $\hat{z}$. Al pasar por una región con un campo magnético inhomogéneo la trayectoria de un pequeño imán se desviará en un ángulo que depende del valor de la proyección del imán en la dirección del campo magnético del aparato. Esto no es difícil de comprender, si el campo externo es más intenso en la parte superior del aparato entonces los imanes que ingresen al mismo con su polo norte apuntando hacia arriba se desviarán en esa misma dirección (moviéndose hacia zonas donde el campo magnético es más intenso. Groseramente, podemos pensar que en ese caso la fuerza que tendería a desplazar al polo norte del imán hacia arriba es mayor que la que empujaría al polo sur hacia abajo). Por el contrario, si un imán ingresa al aparato con su polo norte apuntando hacia abajo tenderá a desviarse en esa misma dirección moviéndose hacia zonas donde el campo magnético es menor (al igual que en el caso anterior podemos razonar groseramente diciendo que la fuerza que empuja el polo sur hacia abajo será mayor que la que tiende a mover el polo norte hacia arriba). En consecuencia, el dispositivo de la Figura logra que las partículas se desvíen de manera diferente según sea la orientación de su spin. Al realizar este experimento por primera vez, Stern y Gerlach observaron un resultado sorprendente: las partículas incidentes (átomos de plata, en ese caso) se desviaban siguiendo solamente dos trayectorias distintas (y que entre estas dos no había un continuo de desviaciones intermedias). Eso indica que, como dijimos más arriba, la proyección del spin en la dirección $\hat{z}$ sólo toma dos valores posibles. A partir de aquí a estos dos valores los llamaremos Z= +1 y Z= -1.[6]

---

[5] Restringiremos nuestra discusión al caso de partículas con "spin 1/2" para las cuales lo anterior es cierto: la medición del spin siempre da lugar a dos resultados; hay otras partículas para las cuales el número de resultados es mayor pero las consideraciones que realizamos aquí se aplican también a ellas, con algunas pequeñas variantes.

[6] En un lenguaje más técnico, los valores de las componentes del spin son siempre múltiplos de la cantidad $\hbar/2$; en esas unidades dichas componentes toman simplemente los valores +1 o -1.

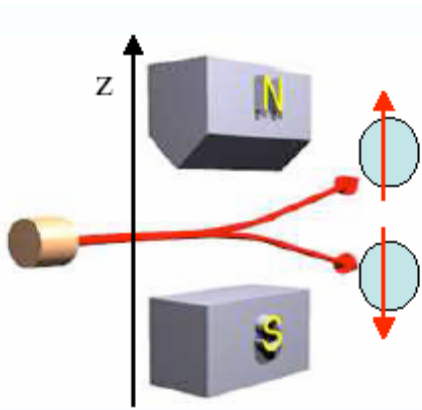

Figura 1: Cuando un haz de partículas con spin 1/2 atraviesa un campo magnético que aumenta en la dirección $\hat{z}$ se divide en dos componentes (una para cada valor de la componente $\hat{z}$ del spin).

La medición de la proyección del spin a lo largo de cualquier dirección puede hacerse rotando el aparato descripto en la Figura 1 (haciendo coincidir la dirección deseada con eje $\hat{z}$ de la Figura). Realizando experimentos como el descripto se observa el mismo fenómeno: siempre las partículas se desvían hacia arriba o hacia abajo. Nunca hay desviaciones intermedias. Por lo tanto nos vemos obligados a concluir que las componentes del spin a lo largo de cualquier dirección toman solamente dos valores. En lo que sigue vamos a utilizar la siguiente notación: las letras minúsculas $\hat{a}$, $\hat{b}$, $\hat{c}$, …, $\hat{x}$, $\hat{y}$, $\hat{z}$ serán usadas para denotar diferentes direcciones del espacio. Las proyecciones del spin a lo largo de cada una de estas direcciones será denotada con la correspondiente letra mayúscula: A, B, C,…, X, Y, Z (o sea, A será la componente del spin a lo largo de la dirección $\hat{a}$). A partir de los experimentos que acabamos de describir debemos concluir que los valores medidos de las propiedades A, B, C,…, X, Y, Z, son siempre +1 o -1.

### D. El spin y la complementariedad

El hecho de que solamente obtengamos dos resultados en la medición de cualquier componente del spin es realmente extraño. En efecto, esto nos dice que el spin no puede describirse con una flecha ordinaria ya que si imaginamos que esa flecha tiene componentes cuyos valores son iguales a +1 o -1 en cualquier dirección llegaríamos inevitablemente a una contradicción. No es difícil convencerse de esto razonando "por el absurdo": por simplicidad imaginemos al spin como una flecha que apunta en alguna dirección del plano del papel y supongamos que sus componentes a lo largo de dos direcciones perpendiculares son iguales a +1. Una flecha como esta se observa en la Figura 2. Examinando esa figura es fácil darse cuenta que, usando el teorema de Pitágoras, la componente de esa flecha a lo largo de una dirección que forma un ángulo de 45º grados con las anteriores tendría un valor igual $\sqrt{2}$. ¡Pero esto es incompatible con los resultados experimentales que nos indican que si medimos la componente del spin a lo largo de esa dirección también obtendremos los valores +1 o -1! Evidentemente, el spin es un vector muy extraño ya que los valores medidos para cada una de sus componentes no nos permiten formarnos una imagen coherente del spin como un vector. ¿Cómo resuelve la mecánica cuántica este problema? Pues bien, lo hace apelando al principio de complementariedad: establece que las componentes del spin a lo largo de tres direcciones perpendiculares son propiedades complementarias, y por lo tanto no pueden ser determinados simultáneamente.

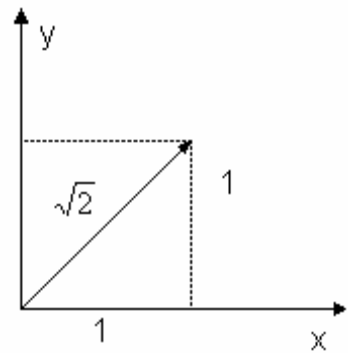

Figura 2: Un vector que tiene componentes iguales a +1 en dos direcciones perpendiculares tiene una componente igual a $\sqrt{2}$ a lo largo de una dirección intermedia. Esto nos muestra que no es posible formarse una imagen intuitiva del spin como un vector ordinario ya que si midiéramos el spin en la dirección intermedia también obtendríamos que su proyección toma los valores +1 o -1.

El carácter complementario de las componentes del spin a lo largo de tres direcciones perpendiculares ($\hat{x}$, $\hat{y}$, $\hat{z}$) se puede poner en evidencia analizando los resultados de secuencias de experimentos como el descripto en la Figura 1. Por ejemplo, supongamos que realizamos primero una medición de la componente $\hat{z}$ del spin tal como describimos más arriba. Utilizamos un aparato como el de la Figura 1 y observamos que las partículas se desvían hacia arriba o hacia abajo. Seguidamente, introducimos las partículas que tienen Z=+1 (aquellas que se desviaron hacia arriba) y las introducimos en un aparato que mide la componente del spin a lo largo de la dirección $\hat{x}$ (perpendicular a $\hat{z}$). ¿Cuál es el resultado? Pues bien, si repetimos muchas veces este experimento podemos comprobar que en la mitad de los casos obtenemos X=+1 (las partículas se desvían hacia la derecha) y en la otra mitad obtenemos X=-1 (las partículas se desvían hacia la izquierda). Por lo tanto, concluimos que la medición de la propiedad X realizada sobre partículas que tienen Z=+1 da lugar a resultados completamente aleatorios. Esto es precisamente lo que muestra el carácter complementario de las propiedades X y Z. Algo análogo ocurre con la propiedad Y (que es complementaria a las dos anteriores).

### E. Entrelazamiento entre las partes de un sistema compuesto

Hasta aquí hemos descripto propiedades sorprendentes de la mecánica cuántica, pero todavía debemos analizar la más sorprendente de todas ellas. En efecto, de acuerdo a la mecánica cuántica la naturaleza de los sistemas compuestos es profundamente anti-intuitiva. La física clásica y el sentido común indican que una descripción completa del sistema compuesto es equivalente a una descripción completa de cada una de sus componentes. Sin embargo esto es falso en el mundo cuántico. Por el contrario, de acuerdo a la mecánica cuántica es posible que un sistema compuesto tenga sus propiedades máximamente determinadas, aunque las propiedades de cada una de sus partes sean completamente aleatorias. ¿Cómo es esto? Consideremos una propiedad de un sistema formado por dos partículas. Para fijar ideas, consideremos la propiedad $X_{12}=X_1 X_2$ para un sistema de dos partículas de spin 1/2. De acuerdo a la mecánica cuántica esta propiedad también puede tomar dos valores que son: $X_{12}=+1$ y $X_{12}=-1$. Esto no es sorprendente ya que $X_{12}$ se define como el producto de $X_1$ y $X_2$. Ambos factores pueden tomar los valores ±1 y por lo tanto el producto de ellos también será igual a +1 o -1. Por ejemplo, podríamos obtener $X_{12}=-1$ como el producto de $X_1=+1$ y $X_2=-1$ o bien como el producto de $X_1=-1$ y $X_2=+1$.

Pero la mecánica cuántica nos dice que las propiedades de un sistema compuesto no necesariamente deben determinarse a partir de mediciones sobre las partes. Por ejemplo, para determinar el valor de la propiedad $X_{12}$ tenemos la opción de determinar por separado los valores de $X_1$ y $X_2$. Pero también tenemos otras opciones, podemos determinar $X_{12}$ sin medir ambos factores por separado. Para eso hay que interactuar directamente con el todo y no con las partes. Esto puede ser difícil en la práctica, sobre todo, cuando el sistema compuesto se fragmenta naturalmente y las partes viajan en direcciones diferentes. Pero la mecánica cuántica nos plantea no solamente que esto es posible sino que si logramos hacerlo podremos observar comportamientos muy extraños. Por ejemplo, podríamos determinar que el valor de la propiedad "colectiva" $X_{12}$ resulta ser $X_{12}=-1$. Naturalmente, si repetimos esta medición verificaríamos que obtenemos siempre el mismo resultado: $X_{12}=-1$. Pero ahora, si realizamos mediciones de $X_1$ en el primer fragmento, observaríamos que los valores obtenidos son totalmente azarosos. Lo mismo ocurriría si midiéramos $X_2$. Pero verificaríamos que si multiplicamos los valores obtenidos en estas mediciones "locales" el resultado sería siempre el mismo: verificaríamos que $X_1 X_2=-1$. Apelando a una metáfora, la mecánica cuántica permite que un objeto tenga una propiedad colectiva con un valor preciso (por ejemplo $X_{12}=-1$) pero que en ese objeto coexistan todavía dos alternativas drásticamente distintas para sus partes: (en este caso esas alternativas son ($X_1=+1$, $X_2=-1$) y ($X_1=-1, X_2=+1$)). Cuando un objeto tiene este comportamiento decimos que se encuentra en un "estado entrelazado" (en inglés se utiliza el nombre "entangled state", que ha dado lugar a más de una traducción al español).

Los estados entrelazados no tienen ninguna contraparte clásica. En estos estados la identidad del todo está perfectamente definida pero la identidad de las partes está máximamente indeterminada. De acuerdo a Erwin Schrödinger, uno de los padres de la física cuántica, la existencia del entrelazamiento es la característica que hace que esta materia sea verdaderamente irreconciliable con la intuición clásica.

Es importante que describamos con más detalle el ejemplo más sencillo de entrelazamiento. Para esto basta con completar el ejemplo que acabamos de presentar. Consideremos dos partículas de spin 1/2. Tal como lo hicimos más arriba, denotamos a las componentes del spin de cada partícula en las direcciones de los ejes $\hat{x}$ e $\hat{y}$ como $X_1$, $X_2$, $Y_1$, $Y_2$. Sabemos que las propiedades $X_1$ e $Y_1$ son complementarias entre sí y, por lo tanto, no pueden ser determinadas simultáneamente. Lo mismo sucede con las propiedades $X_2$ e $Y_2$. Sin embargo, la mecánica cuántica permite la existencia de estados donde las propiedades

$$X_{12} = X_1 X_2$$
$$Y_{12} = Y_1 Y_2 \qquad (1)$$

están bien determinadas simultáneamente. Es decir, pese a la complementariedad de las propiedades X e Y para cada partícula, las propiedades colectivas $X_{12}$ e $Y_{12}$ NO son complementarias (aunque estén construidas a partir de ingredientes complementarios). Es posible diseñar y construir un aparato de medición que determine simultáneamente el valor de las propiedades $X_{12}$ e $Y_{12}$ (omitiremos aquí toda discusión técnica al respecto, baste decir que para que este aparato funcione es necesario controlar las interacciones entre ambas partículas de manera muy sutil).

Supongamos que determinamos simultáneamente $X_{12}$ e $Y_{12}$ y obtenemos que dichas propiedades toman valores $X_{12}$=-1 e $Y_{12}$=-1. Como dijimos más arriba, si realizamos mediciones locales (separadamente medimos las componentes del spin de cada partícula) verificaremos que los resultados obtenidos son aleatorios. Pero observaremos que los resultados de los experimentos sobre las partes están fuertemente correlacionados. Por ejemplo, observaremos que los productos de los resultados siempre son negativos: $X_1 X_2$=-1 e $Y_1 Y_2$=-1. Estas correlaciones son extremadamente fuertes y, como veremos, pueden utilizarse para predecir comportamientos que entran en violenta contradicción con la intuición clásica.

## III. El ataque final de Einstein contra la mecánica cuántica

En 1935 Einstein, Podolsky y Rosen (EPR) publicaron en el *Physical Review* un artículo con un título provocativo en el que preguntaban: "*¿Puede considerarse que la descripción cuántica de la realidad física es completa?*". En el trabajo argumentaban que la respuesta a esta pregunta es negativa. Einstein creía haber encontrado un argumento que permitía demostrar que en la mecánica cuántica anidaba el germen de su propia destrucción. Sin embargo, como veremos, su profecía resultó ser incorrecta (hoy sabemos con certeza que si este germen existe, no es aquel encontrado por Einstein en 1935).

En su célebre trabajo, EPR establecen en primer lugar una serie de requisitos que toda teoría que aspire a describir la realidad física debe cumplir. De acuerdo a los autores, las teorías físicas tienen que tener a los "elementos de la realidad" como principales protagonistas. EPR proveen una definición operacional para distinguir aquellas propiedades de la naturaleza que deben ser considerados "elementos de la realidad". Esta definición es la siguiente: *Si somos capaces de predecir con certeza el valor de alguna propiedad de un objeto sin perturbarlo en modo alguno, entonces esa propiedad debe ser considerada un "elemento de la realidad"*. La idea es simple: si nuestra predicción no afecta en modo alguno al sistema, la propiedad en cuestión tiene que tener un sustrato real, su valor debe de estar "escrito" en el objeto en cuestión. Estos criterios propuestos por EPR para toda teoría física pueden ser discutidos en el plano epistemológico o filosófico, pero suenan aceptables para la mayoría de las personas. El objetivo del trabajo de EPR es demostrar que la mecánica cuántica no cumple con estos principios y que, por lo tanto, no puede ser considerada una descripción completa de la realidad física.

La clave del trabajo de EPR consiste en analizar las propiedades de los estados entrelazados. El nudo del argumento EPR (en la versión desarrollada más tarde por David Bohm) es el siguiente: consideremos un sistema compuesto por dos partículas de spin 1/2 que es preparado de modo tal que sus propiedades $X_{12}$ e $Y_{12}$ (definidas más arriba) toman los valores $X_{12}=-1$ y $Y_{12}=-1$. Consideremos además que las partículas 1 y 2 pueden ser separadas y llevadas a laboratorios distantes que llamaremos Labo-1 y Labo-2. Utilizaremos laboratorios tan separados como queramos, como para que ninguna perturbación material generada durante las mediciones realizadas en el Labo-1 tenga tiempo suficiente para propagarse hasta el Labo-2 (y viceversa). Tal como discutimos más arriba, si en el Labo-1 medimos la propiedad $X_1$ sobre la primera partícula podemos predecir el resultado que obtendríamos si midiéramos $X_2$ en el Labo-2. En efecto, sabemos que si obtenemos $X_1=+1$ entonces con certeza podemos predecir que si midiéramos $X_2$ deberíamos obtener el resultado $X_2=-1$. Análogamente, si obtenemos $X_1=-1$ entonces predecimos con certeza que si decidiéramos medir $X_2$ obtendremos el valor $X_2=+1$. Por lo tanto el valor de la propiedad $X_2$ siempre puede ser predicha con certeza a partir de los resultados de experiencias realizadas en el Labo-1, que es un laboratorio tan distante que ninguno de los eventos que ocurren en su interior puede alterar el estado de las cosas para la partícula 2.

En consecuencia, estamos obligados a concluir que $X_2$ debe ser un "elemento de la realidad". Lo mismo debe pasar con $Y_2$ ya que podríamos predecir con certeza su valor a partir de experimentos del mismo tipo, que involucran medir la propiedad $Y_1$ sobre la primera partícula. La conclusión a la que nos conduce este razonamiento es que tanto $X_2$ como $Y_2$ son "elementos de la realidad" y por lo tanto tienen que tener un lugar dentro de una teoría física completa. Sin embargo, para la mecánica cuántica estas propiedades son complementarias y sus valores no pueden ser definidos simultáneamente. En consecuencia, concluyen EPR: la mecánica cuántica no puede proveer una descripción completa de la realidad física.

El trabajo de EPR recibió una rápida (y breve) respuesta de Niels Bohr quien hizo notar que el argumento de EPR no expone en realidad ninguna contradicción interna de la mecánica cuántica. Por otra parte Bohr destacó que el argumento de EPR utiliza un razonamiento "contra-fáctico" ya que mezcla resultados de experimentos reales con resultados de experimentos imaginarios. En efecto: en el primer laboratorio tenemos que decidir que propiedad mediremos para la partícula 1. Podríamos elegir medir $X_1$ o bien podríamos elegir medir $Y_1$. Pero no podemos hacer las dos cosas a la vez. El argumento EPR mezcla sutilmente los resultados de ambas mediciones ya que en definitiva ambas son necesarias si pretendemos otorgar el status de "elementos de realidad" tanto a la propiedad $X_2$ como a $Y_2$. Efectivamente, aquí hay un razonamiento contra-fáctico. Pero es un razonamiento que cualquier persona sensata estaría dispuesta a hacer. Si la partícula 2 se encuentra en el Labo-2, nada puede saber sobre cuál es la propiedad que el experimentador decidirá medir en el Labo-1. En consecuencia, deberíamos estar dispuestos a aceptar que, pese a que no podemos realizar los dos experimentos sino que debemos elegir uno de ellos, tanto las propiedades $X_2$ como $Y_2$ deben estar escritas en la segunda partícula (o sea, deben ser "elementos de la realidad"). En cambio, la mecánica cuántica no nos permite razonar de esta forma. Asher Peres acuñó una frase que describe la actitud que debería tener un físico pragmático ante la posibilidad de caer en razonamientos contra fácticos. No debería olvidar nunca que *los experimentos que no se realizan no tienen resultados*.

### A. Las variables ocultas y las teorías realistas-locales, ¿una escapatoria?

Queda claro que el argumento de EPR no demuestra una inconsistencia interna de la mecánica cuántica sino que pone en evidencia que esta teoría no satisface ciertos criterios de muy razonable apariencia. Naturalmente debemos preguntarnos si es posible que exista una alternativa compatible con los resultados de los experimentos (que hasta el día de hoy coinciden con las predicciones de la mecánica cuántica) y que además sea compatible con el sentido común o, más precisamente, con los postulados de EPR. Una teoría de estas características fue mencionada más arriba. Podríamos imaginar que existen en la naturaleza grados de libertad microscópicos que todavía no hemos sido capaces de descubrir. Estos grados de libertad son usualmente denominados "variables ocultas". Si existieran variables ocultas, podríamos concebir la posibilidad de que nuestra ignorancia sobre su comportamiento y su naturaleza es la responsable de la aleatoriedad que observamos en los resultados de ciertos experimentos. Es decir, podríamos concebir la posibilidad de que al repetir muchas veces el mismo experimento sin controlar el comportamiento de las variables ocultas estuviéramos generando sistemas que en realidad no son idénticos entre sí. En cada realización experimental, en cada evento, los resultados de los experimentos estarían completamente determinados por los valores ocultos. Pero al repetir muchas veces el mismo experimento podríamos obtener resultados distintos distribuidos de manera aparentemente aleatoria. Esta aleatoriedad sería simplemente una consecuencia de nuestra ignorancia.

El trabajo de EPR tuvo la virtud de exponer de manera sistemática cuales son las propiedades que nuestro sentido común le reclama a las teorías físicas. Las teorías compatibles con el sentido común son aquellas que se engloban con el nombre de *teorías realistas locales*. Diremos que una teoría es "realista" (una palabra que tal vez tiene connotaciones demasiado fuertes como para ser utilizada aquí) si acepta el hecho de que todas las propiedades observables (los elementos de realidad) de los sistemas físicos tienen valores precisos que en última instancia determinan los resultados de las mediciones que efectuemos sobre ellas. Estas teorías incluyen a las que aceptan la existencia de variables ocultas. De acuerdo a ellas la realidad física se describe en su nivel más profundo mediante un modelo en el que los resultados de todos los posibles experimentos están escritos de algún modo en los objetos. Es decir, en este contexto el realismo es sinónimo de determinismo. Toda aleatoriedad debe originarse en nuestra limitada capacidad de control o de conocimiento. Diremos que una teoría es "local" si no admite la posibilidad de que exista acción a distancia o propagación instantánea de cualquier tipo de señal o perturbación. En estas teorías, separando suficientemente dos partes de un sistema (llevándolas a laboratorios muy distantes) garantizamos que las acciones que realicemos en un laboratorio no tendrán ninguna influencia sobre lo que suceda en el otro laboratorio.

La posibilidad de que exista alguna teoría más fundamental que la mecánica cuántica basada en variables ocultas fue considerada por numerosos autores. La discusión sobre este asunto se aplacó luego de que John von Neumann publicara un teorema en el que se demostraba que no era posible construir una teoría de este tipo que diera lugar a las mismas predicciones que la mecánica cuántica. Sin embargo a principios de los años 60, John Bell puntualizó que el teorema de von Neumann contenía un error, una hipótesis demasiado restrictiva que hacía que sus consecuencias no fueran trascendentes. El propio Bell, comenzó a explorar entonces la posibilidad de construir teorías de variables ocultas dando lugar a una serie de trabajos de consecuencias notables.

## IV. Desigualdades de Bell: mecánica cuántica contra teorías realistas-locales

Los trabajos de John Bell permitieron que la discusión sobre la existencia de teorías de variables ocultas pasara del terreno de la filosofía al de la física, en el cual la validez de los modelos es sometidos al juicio de los experimentos. Es interesante notar que la intención de John Bell al comenzar sus investigaciones era encontrar argumentos a favor del punto de vista de Einstein. Bell expuso su posición ideológica con elocuencia: *"Yo pensaba que la superioridad intelectual de Einstein sobre Bohr en este punto era enorme: una distancia gigante entre un hombre que veía claramente lo que se necesitaba (Einstein) y un oscurantista (Bohr)"*. Paradójicamente, con sus trabajos Bell logró exactamente lo contrario de lo que se proponía: descubrió la forma en la cual el punto de vista de Einstein podía demostrarse falso a partir de los resultados de experimentos reales.

La trascendencia de los trabajos de Bell no puede subestimarse. Los mismos han tenido un impacto enorme en las últimas décadas. En breves palabras, Bell demostró que todas las teorías realistas locales conducen a predicciones cuantitativas sobre resultados experimentales concretos. Asimismo, demostró que estas predicciones pueden entrar en contradicción con las de la mecánica cuántica. En consecuencia, la validez de uno u otro modelo (el cuántico o aquel basado en nuestro sentido común) puede ser sometida al juicio de la ciencia experimental.

A primera vista resulta sorprendente que sea posible derivar predicciones para todas las teorías realistas locales. Estas predicciones toman la forma de desigualdades matemáticas que restringen los valores que pueden tomar las probabilidades de eventos registrados en laboratorios distantes cuando se realizan experimentos sobre las partes de un sistema compuesto. Estas relaciones matemáticas se conocen con el nombre de "desigualdades de Bell". En lo que sigue presentaremos una deducción sencilla de una de estas desigualdades (que no fue presentada por Bell sino por David Mermin en 1981).

## A. Un experimento sencillo realizado en dos laboratorios

Consideremos ahora una situación como la analizada en el trabajo de EPR (en la versión desarrollada por David Bohm). Tomamos un sistema compuesto por dos partículas de spin 1/2. Determinamos simultáneamente los valores de las propiedades $X_{12}=-1$ e $Y_{12}=-1$, creando de este modo un estado entrelazado cuyas propiedades discutimos más arriba. Luego llevamos a cada partícula a un laboratorio distinto (Labo-1 y Labo-2). Ambos laboratorios están espacialmente separados y la distancia entre ellos es tal que no hay posibilidad de propagación de ninguna señal de un laboratorio a otro durante el tiempo en que transcurren nuestros experimentos. En cada laboratorio un experimentador medirá la componente del spin de su partícula a lo largo de alguna de las tres direcciones que indicamos como â, $\hat{b}$ o $\hat{c}$ en la Figura 3 (las tres direcciones forman un ángulo de 120º grados entre sí). Los experimentadores que actúan en cada uno de sus laboratorios eligen al azar en cual de las tres direcciones miden el spin. Podemos pensar que cada experimentador tiene a su disposición un aparato como el que aparece en la Figura 3. Dicho aparato tiene un selector con tres posiciones. Cuando el selector apunta hacia la izquierda el aparato mide la componente â del spin, si el selector apunta hacia arriba el aparato mide la componente $\hat{b}$ y si apunta hacia la derecha mide la componente $\hat{c}$. Cualquiera de esas mediciones da lugar solamente a dos resultados: +1 o -1. El experimento se repite muchas veces y en cada repetición el sistema se prepara de manera idéntica, ambas partículas se separan y cada experimentador elige al azar (y de manera totalmente independiente) la posición del selector de su aparato y registra el valor que obtiene en su medición.

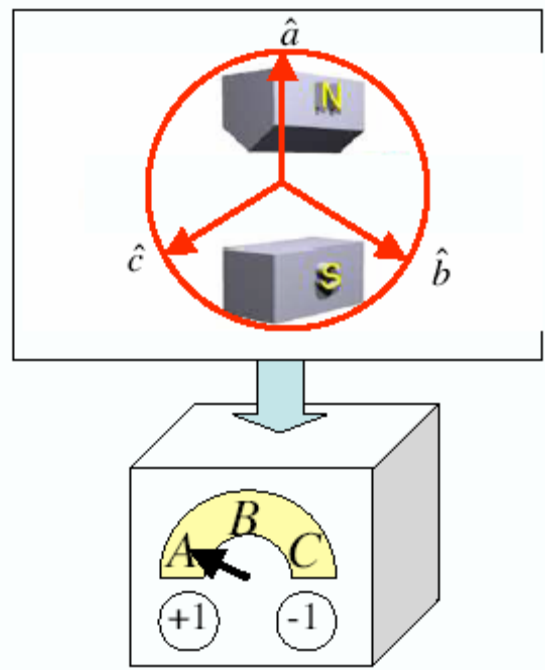

Figura 3: para poner a prueba la versión más sencilla de las desigualdades de Bell es necesario un aparato que mide el valor de la componente â, $\hat{b}$ o $\hat{c}$ de una partícula de spin 1/2.

Como cada experimentador puede elegir medir una de tres propiedades (A, B, o C) las mediciones realizadas en los dos laboratorios se pueden agrupar en nueve configuraciones. Sin mucho esfuerzo podemos hacer una lista de todas ellas. Colocando en primer lugar la propiedad medida en el Labo-1 y en segundo lugar la que se mide en el Labo-2, las nueve configuraciones son: $A_1$- $A_2$, $B_1$- $B_2$, $C_1$- $C_2$, $A_1$- $B_2$, $B_1$- $A_2$, $A_1$- $C_2$, $C_1$- $A_2$, $B_1$- $C_2$ y $C_1$- $B_2$.

## B. El experimento según las teorías realistas locales

Pensemos de qué manera describiría esta situación experimental una teoría realista local. En primer lugar, cualquier teoría de este tipo debe aceptar que antes de que el experimentador que trabaja en el Labo-1 decida que propiedad medirá, el resultado de dicha medición debe tener existencia real en la partícula 1. Esto es así porque las tres propiedades que el experimentador puede medir (que denotamos como A, B o C) son "elementos de la realidad". En efecto, el argumento EPR debería ser suficiente para convencernos de esto: los valores de estas propiedades podrían ser predichos con certeza si hiciéramos el experimento adecuado en el Labo-2. Entonces, todas las teorías realistas locales deben aceptar que cada partícula viaja hacia su detector llevando un conjunto de instrucciones consigo que indican el resultado de cualquier medición. Es tentador utilizar una metáfora biológica: Cada partícula lleva consigo *genes* que determinan los valores de las propiedades A, B, C. Podríamos denotar a estas instrucciones con una terna de números (A,B,C) que indican los valores que se obtendrían si se midiera el valor de alguna de estas tres propiedades. Por ejemplo si la partícula lleva un gen del tipo (+1,-1,+1) quiere decir que si el experimentador decidiera medir A o C obtendría en ambos casos el valor +1 mientras que si midiera B el resultado sería -1. Es evidente que, como solamente hay dos resultados posibles para la medición de cada una de las tres propiedades, tan solo hay ocho genes posibles para cada partícula. En la siguiente Tabla presentamos la lista exhaustiva de todos ellos:

| Genes posibles | |
|---|---|
| Partícula 1 | Partícula 2 |
| (+1,+1,+1) | (-1,-1,-1) |
| (+1,+1,-1) | (-1,-1,+1) |
| (+1,-1,+1) | (-1,+1,-1) |
| (+1,-1,-1) | (-1,+1,+1) |
| (-1,+1,+1) | (+1,-1,-1) |
| (-1,+1,-1) | (+1,-1,+1) |
| (-1,-1,+1) | (+1,+1,-1) |
| (-1,-1,-1) | (+1,+1,+1) |

Por otra parte, toda teoría realista local debe aceptar que los genes que lleva la partícula 1 tienen que estar correlacionados con los que lleva la partícula 2. En efecto, esto debe ser así porque si los dos experimentadores decidieran medir la misma propiedad verificarían que obtienen resultados opuestos. Por lo tanto, el gen que lleva la primera partícula determina completamente al gen de la segunda. Por ejemplo, si la primera partícula lleva un gen del tipo (+1,-1,+1) la segunda debe llevar un gen complementario, del tipo (-1,+1,-1).

## C. La desigualdad de Bell más sencilla

El descubrimiento fundamental de Bell es que todas las teorías que aceptan la existencia de genes deben satisfacer ciertas restricciones, que toman la forma de desigualdades matemáticas. Presentaremos aquí la versión más sencilla de estas desigualdades. Invitamos al lector a realizar un intento por seguir el siguiente razonamiento, que resultará crucial para el resto de nuestro argumento.

Supongamos que la primera partícula lleva el gen (+1,+1,+1). En ese caso, la segunda llevará el gen (-1,-1,-1). Entonces, aunque los dos experimentadores midan propiedades distintas los resultados que obtendrán serán siempre opuestos: en el Labo-1 siempre se obtendrá el resultado +1 mientras que en el Labo-2 siempre se obtendrá el resultado -1. Una situación idéntica tiene lugar si el gen que lleva la primera partícula es (-1,-1,-1) ya que en ese caso también los resultados serán siempre opuestos. Si las partículas fueran generadas únicamente con estos dos tipos de genes entonces deberíamos concluir que los resultados obtenidos en ambos laboratorios serían siempre opuestos. Por supuesto, esta no es una hipótesis razonable ya que no sabemos nada sobre el mecanismo subyacente que produce genes diferentes (esas son, precisamente, las variables ocultas).

Pero, aunque parezca mentira, es posible deducir una propiedad muy sencilla que se debe cumplir para todos los otros genes (o sea, aquellos en los que hay una instrucción que es distinta de las otras dos como es el caso de los genes (+1,+1,-1) y (+1,-1,+1)). Es fácil mostrar que para todos esos genes *habrá cinco configuraciones para las cuales los resultados obtenidos en Labo-1 y Labo-2 serán distintos y cuatro configuraciones para las cuales estos resultados serán iguales*. Para ver que esto es cierto es suficiente con hacer un análisis exhaustivo de lo que sucede con cada uno de los genes. Por ejemplo, si el gen que lleva la primera partícula es (+1,+1,-1), tal como está indicado en la Figura 4, los resultados de los experimentos serán opuestos siempre que el primer y segundo experimentador midan respectivamente las propiedades $A_1$- $A_2$, $A_1$- $B_2$, $B_1$- $B_2$, $B_1$- $A_2$ y $C_1$- $C_2$. En cambio, los resultados serán idénticos siempre que los experimentadores realicen las mediciones de las propiedades $A_1$- $C_2$, $C_1$- $A_2$, $B_1$- $C_2$ y $C_1$- $B_2$. El lector puede comprobar que para todos los genes en los que las tres instrucciones no sean idénticas se verifica este mismo resultado: *siempre hay cinco configuraciones de los detectores para los que los resultados obtenidos en ambos laboratorios son opuestos y hay cuatro para las cuales los resultados son idénticos* (recordemos que si las instrucciones son idénticas entonces los resultados siempre serán distintos). Si los experimentadores eligen al azar las configuraciones de sus detectores, entonces podemos concluir que por lo menos en 5 de cada 9 experimentos los resultados serán opuestos.

| GEN DE LA PARTICULA 1: (+1,+1,-1) | | | |
|---|---|---|---|
| Cinco experimentos con resultados distintos | | Cuatro experimentos con resultados iguales | |
| Labo-1 | Labo-2 | Labo-1 | Labo-2 |
| A | A | A | C |
| B | B | C | A |
| C | C | B | C |
| A | B | C | B |
| B | A | | |

Figura 4: Para el gen (+1,+1,-1) hay cinco configuraciones de los detectores que dan lugar a que el resultado registrado en el Labo-1 sea diferente que el registrado en el Labo-2 mientras que hay cuatro configuraciones para las cuales los resultados son idénticos. Esto se repite para todos los genes en los que las tres instrucciones no son iguales.

Esta predicción es totalmente independiente de la naturaleza de las variables ocultas. Esta conclusión es tan importante que merece ser repetida. Para toda teoría realista local predecimos que la probabilidad $P_{R-L}$ de que se obtengan resultados diferentes debe cumplir la siguiente desigualdad:

$P_{R-L}$(Labo-1 $\neq$ Labo-2)$\geq$5/9=0.5555… (2)

### D. El experimento según la mecánica cuántica

La mecánica cuántica también realiza una predicción para el valor de la probabilidad de que se obtengan resultados diferentes en ambos laboratorios. Esta predicción es drásticamente diferente de la de las teorías realistas locales. En efecto, de acuerdo a la mecánica cuántica la probabilidad de obtener resultados distintos es:

$P_{Cuant}$(Labo-1 $\neq$ Labo-2)=1/2=0.5 (3)

Para llegar a esta conclusión es necesario utilizar el formalismo matemático de la mecánica cuántica. Sin embargo, podemos hacer un intento por explicar su origen de manera sencilla (el lector no interesado está invitado a omitir la lectura de este párrafo). Si realizamos mediciones sucesivas de componentes de un spin de una partícula en direcciones perpendiculares sabemos que, como las proyecciones perpendiculares del spin definen magnitudes complementarias, los resultados de la segunda medición estarán distribuidos al azar con una probabilidad del 50% para cada uno de los dos valores posibles. En cambio, si realizamos mediciones sucesivas en dos direcciones $\hat{a}$ y $\hat{b}$, que forman un ángulo $\theta_{\hat{a}\hat{b}}$, la mecánica cuántica establece que la probabilidad de que los resultados de ambas mediciones serán iguales es

$$P(B = A) = \cos^2(\theta_{\hat{a}\hat{b}}/2)$$

Si las direcciones $\hat{a}$ y $\hat{b}$ forman un ángulo de 120º grados, como en el caso de la Figura 4, la probabilidad de que los resultados de dos mediciones sucesivas sean iguales es 1/4 (o sea, en el 25% de los casos obtendremos resultados iguales y en el 75% obtendremos resultados distintos[7]). Con este ingrediente estamos en condiciones de deducir cual es la predicción que la mecánica cuántica realiza para el experimento analizado en las secciones anteriores. Para calcular la probabilidad de que los resultados del Labo-1 sean diferentes de los del Labo-2 tenemos que analizar todos los casos posibles. Presentaremos aquí el estudio de uno de ellos y dejaremos para el lector interesado el examen del resto, que se realiza con un razonamiento similar. Supongamos que en el Labo-1 se midió la propiedad A y se obtuvo el valor +1. En ese caso sabemos que si midiéramos la propiedad A en el Labo-2 obtendríamos con certeza el valor (-1). En consecuencia podemos afirmar que la partícula que se encuentra en el Labo-2 está en el estado de spin -1 en la dirección $\hat{a}$. Nos interesa calcular en ese caso cual es la probabilidad de obtener el valor -1 para la medición de las componentes $\hat{a}$, $\hat{b}$ o $\hat{c}$. Para eso podemos analizar todos los casos posibles. Si medimos A (lo que ocurre en la tercera parte de los casos) obtendremos el resultado -1 con probabilidad 1. En cambio, si medimos B o C (lo que ocurre en las restantes dos terceras partes de los casos) podemos apelar al resultado que mencionamos más arriba y afirmar que obtendremos el valor -1 con probabilidad 1/4. En conclusión si en el Labo-1 se mide A=+1 la probabilidad de que los resultados de las mediciones realizadas en el Labo-2 sean distintas resulta ser (1+1/4+1/4)/3=1/2, que es justamente el resultado que mencionamos más arriba. Razonando de igual modo para los restantes resultados posibles para las mediciones realizadas en el Labo-1 llegamos a la misma conclusión y de ese modo demostramos la validez de la predicción cuántica expresada más arriba.

El contraste entre la predicción cuántica y la predicción de cualquier teoría realista-local es drástico: de acuerdo a la mecánica cuántica en la mitad de los experimentos obtendremos resultados diferentes y en la otra mitad los resultados serán idénticos. Esto es incompatible con la predicción de cualquier teoría realista local ya que de acuerdo a todas ellas los resultados deben ser diferentes por lo menos en el 55.5% de los experimentos. ¿Quién tiene razón, la mecánica cuántica o las teorías realistas locales? Para dirimir este debate, debemos realizar el experimento y comprobar cual de las dos predicciones es la correcta.

## V. La violación de las desigualdades de Bell

Después de los trabajos de Bell varios grupos se lanzaron a realizar experimentos como los descriptos en la sección anterior. Cabe aclarar que ninguno de estos grupos lo hizo con la esperanza de detectar violaciones a las predicciones cuánticas. Por el contrario, a esa altura del siglo XX nadie dudaba que la mecánica cuántica saldría airosa en su confrontación contra las teorías de variables ocultas. Para poder realizar estos experimentos fue necesario superar varios obstáculos tecnológicos y los primeros resultados en los que se detectaron claras violaciones a las desigualdades de Bell fueron obtenidos recién en 1982 por Alain Aspect en Paris.

El experimento de Aspect fue un verdadero *tour de force* por el que debería hacerse acreedor al premio Nóbel de física. Fue realizado utilizando pares de fotones entrelazados generados a partir del decaimiento de átomos de Calcio. Cuando este tipo de átomo decae en una cascada S-P-S emite dos fotones que tienen casi la misma frecuencia y que están entrelazados en su polarización. Este grado de libertad de los fotones se comporta de manera muy similar al spin de una partícula de spin 1/2. Para realizar su experimento Aspect no solamente tuvo que perfeccionar su fuente de pares de fotones entrelazados (que para esa época eran toda una novedad). Una vez producidos cada uno de los fotones se dirigía hacia un extremo distinto del laboratorio donde se habían montado dos estaciones de trabajo idénticas que jugaban el rol del Labo-1 y el Labo-2 que mencionamos más arriba. Estas estaciones constaban de un detector que cumplía el papel del instrumento de medición que ilustramos en la Figura 3. En el experimento, en cada estación de trabajo los fotones se encontraban con un espejo que cambiaba de orientación de manera azarosa. Para cada una de estas direcciones los fotones eran enviados a detectores diferentes en los que se medía la polarización en alguna dirección (las que juegan un papel equivalente a las direcciones $\hat{a}$, $\hat{b}$ o $\hat{c}$ de la Figura 3). Aspect invirtió un esfuerzo considerable para asegurarse de que los espejos variaran su orientación suficientemente rápido y que los detectores estuvieran suficientemente separados como para poder garantizar que no existía conexión causal posible entre los registros tomados en ambos extremos del laboratorio. La longitud del laboratorio era de alrededor de 10 metros y los espejos cambiaban de posición en tiempos del orden de varios nano-segundos (hay que recordar que la luz recorre una distancia de casi treinta centímetros en un nano-segundo).

---
[7] Esto se debe a que el coseno de un ángulo de 60º es igual a 1/2

Los resultados de los experimentos de Aspect fueron concluyentes para la mayoría de los físicos, que por otra parte no dudaban sobre la validez de la mecánica cuántica. Sin embargo, un núcleo de escépticos continuó intentando producir experimentos todavía más concluyentes. Para ellos, los resultados de Aspect podían ser criticados desde distintos ángulos. Por cierto, teniendo en cuenta las implicancias fundamentales del resultado del experimento, se justifica tener una actitud que en otro contexto podría ser calificada de exageradamente conservadora. Los problemas del experimento de Aspect eran fundamentalmente dos. Por un lado los ángulos de los espejos no variaban de manera totalmente aleatoria y por lo tanto era posible imaginar algún mecanismo (inverosímil pero imaginable) por el cual los fotones pudieran "conspirar" para que el experimento pareciera favorecer a la mecánica cuántica aún cuando la teoría subyacente fuera realista local. Por otra parte, el tiempo de respuesta de los detectores era demasiado largo lo cual traía aparejadas limitaciones en la sincronización de eventos (el tiempo de respuesta y el tiempo característico de la emisión en cascada era comparable). Por otra parte, la baja eficiencia de los detectores originaba otro problema potencial: no todos los eventos son registrados y no hay manera de garantizar que el subconjunto de eventos que dan lugar a la señal medida sea una muestra no-sesgada del total. Si bien parece completamente razonable aceptar que esto es cierto, en el contexto de este experimento aún este tipo de suposiciones "razonables" son puestas en discusión. Debido a esta, y a muchos otros cuestionamientos más técnicos, durante las últimas dos décadas del siglo XX se realizaron muchos otros experimentos para testear la violación de las desigualdades de Bell.

En la actualidad las técnicas disponibles para generar pares de fotones entrelazados han avanzado notablemente. Los métodos más modernos utilizan un fenómeno que se conoce como *conversión paramétrica inversa*. Este fenómeno se observa cuando ciertos cristales son iluminados con un láser intenso. Para ciertos cristales no-lineales se produce el proceso de conversión de un fotón del láser en un par de fotones que tienen frecuencias cercanas (en este proceso se conserva la energía y, por lo tanto, la suma de las frecuencias de los fotones emitidos es igual a la frecuencia del láser incidente). El par de fotones resulta estar entrelazado en su polarización. Los fotones generados de este modo han sido utilizados para realizar un gran número de experimentos en los que se demuestra la violación de desigualdades de Bell. Los experimentos actuales involucran distancias mucho mayores que las usadas en el experimento de Aspect. En 2001 el grupo dirigido por Anton Zeillinger en Innsbruck presentó resultados de un notable experimento donde se detectaban violaciones a las desigualdades de Bell con fotones que recorrían varios centenares de metros antes de ser detectados). Poco después, Nicolas Gisin detectó señales claras de violaciones a las desigualdades de Bell en experimentos donde los fotones viajaban decenas de kilómetros (desplazándose por fibras ópticas que corren bajo la superficie del lago de Ginebra). En la actualidad, la existencia de violaciones a las desigualdades de Bell es un hecho que goza de un abrumador consenso a partir de la acumulación de una enorme cantidad de resultados experimentales.

## VI. El entrelazamiento como un recurso físico

El entrelazamiento es una propiedad de la mecánica cuántica que fue reconocida desde sus primeros años. Por ejemplo, es bien sabido que para construir un modelo razonable del átomo de Helio es necesario aceptar que los spines de sus dos electrones están entrelazados. En efecto, los estados entrelazados en sistemas de dos spines surgen muy naturalmente y juegan un rol muy importante en muchos fenómenos de la física atómica y molecular. Ningún físico medianamente informado consideraría al entrelazamiento como una propiedad exótica de la física cuántica. Sin embargo, el tipo de entrelazamiento al que la mayoría de los físicos está acostumbrada es aquel que se produce entre las partes de sistemas microscópicos. En ese contexto las consecuencias paradojales de este fenómeno no se ponen de manifiesto. Pero es evidente que, tal como fue analizado en el trabajo de EPR, cuando el entrelazamiento está presente a escala macroscópica es responsable de buena parte de los misterios de la física cuántica.

Desde hace mucho tiempo que somos concientes de la utilidad de almacenar energía, por ejemplo en una batería. Una vez almacenada es posible utilizarla para prender una lámpara, mover un motor, etc. En definitiva, sabemos que la energía almacenada es útil para realizar trabajo. Sólo recientemente se llegó a la conclusión de que es posible concebir al entrelazamiento como un recurso físico. La pregunta que surge en este contexto es *¿cuál es el tipo de tareas que necesitan del entrelazamiento para su ejecución?* Sólo recientemente se comenzó a abordar esta pregunta y se demostró claramente que, al igual que la energía, podríamos almacenar este recurso y utilizarlo para realizar tareas vinculadas con el procesamiento y la transmisión de la información. La exploración de las posibilidades que abre el uso del entrelazamiento como recurso físico es un campo relativamente nuevo y la demora en su desarrollo se debe a que sólo recientemente se comprobó que es posible generar, preservar y manipular pares de objetos entrelazados sobre distancias macroscópicas.

### A. Teleportación

Por el momento se conocen tan sólo unas pocas tareas que requieren del entrelazamiento para ser completadas. Una de ellas es la *teleportación*. Este nombre fantástico se utiliza para denominar a una tarea que tiene un objetivo mucho más modesto que aquel procedimiento que aparece en muchas series de ciencia ficción (*Star trek,* entre ellas). La teleportación es un proceso mediante el cual el estado de un sistema es trasladado de un laboratorio a otro distante. Para realizar esta tarea es necesario contar con un par de objetos entrelazados que tengan la misma constitución material que el objeto a teleportar (o sea, si queremos teleportar un átomo necesitamos de un par de átomos idénticos en un estado entrelazado). El protocolo de la teleportación, desarrollado en 1993, es el siguiente. Uno de los miembros del par de objetos entrelazados (objeto 1) se transporta al laboratorio de destino y el otro de los integrantes del par (objeto 2) se mantiene en el laboratorio donde está el objeto a teleportar (objeto 3). Es importante destacar que este primer paso involucra el movimiento (transporte) de materia desde el punto donde se encuentra el objeto 3 (el objeto a teleportar) hasta el lugar de destino: la teleportación requiere el transporte de materia. Sin embargo, la materia que se transporta se encuentra en un estado que nada tiene que ver con el estado del objeto a teleportar. El segundo paso del procedimiento consiste en medir un conjunto de propiedades colectivas del objeto a teleportar (objeto 3) y el miembro del par que quedó en su mismo laboratorio (objeto 2). Los valores de las propiedades conjuntas se registran mediante instrumentos convenientemente construidos (omitimos todos los detalles técnicos al respecto) y son transmitidos por un canal ordinario hasta el lugar de destino en el que, como mencionamos antes, se encuentra el segundo miembro del par original (el objeto 1). En resumen, el segundo paso del protocolo de teleportación involucra una medición de una propiedad colectiva y la transmisión de información clásica (el resultado de la medición). Finalmente, el último paso consiste en aplicar una acción física concreta sobre el objeto 1 que depende del resultado de las mediciones efectuadas en el segundo paso (o sea, para cada resultado posible realizamos una acción física posible que puede ser, por ejemplo, mover el objeto una cierta distancia o modificar su momento en una cierta magnitud que dependen del resultado de la medición realizada). La mecánica cuántica garantiza que después de realizar esta acción, el objeto 1 quedará preparado en el mismo estado en el que se encontraba el objeto a teleportar.

Las primeras experiencias de teleportación se realizaron con fotones en el laboratorio de Anton Zeillinger en Innsbruck. El uso de fotones entrelazados presenta una serie de ventajas ya que los métodos de producción de pares se han perfeccionado notablemente. Sin embargo, por razones técnicas, la teleportación con fotones y técnicas ópticas lineales solamente permite alcanzar una eficiencia del 75%. Los experimentos más resonantes con estos métodos fueron realizados por el grupo de Zeillinger que en el año 2004 logró teleportar estados de un fotón entre las dos riveras del río Danubio utilizando fibras ópticas instaladas en el sistema de cloacas de la elegantísima Viena. El grupo de Gisin también alcanzó resultados notables sobre distancias de varios kilómetros recorridos bajo la superficie del lago de Ginebra. La teleportación de estados de átomos fue lograda en notables experimentos realizados por los grupos de David Wineland (en Boulder, EEUU) y de Rainer Blatt (en Innsbruck). En estos casos ambos grupos utilizaron trampas de iones. Esta tecnología permite manipular con mucha precisión el estado interno de átomos ionizados y al mismo tiempo controlar su ubicación y movimiento. Es una tecnología ideal para implementar experimentos como los de teleportación pero por el momento no permite alcanzar separaciones macroscópicas. En efecto, los experimentos de teleportación en trampas de iones han alcanzado distancias de unas pocas decenas de micrones.

### B. Computación cuántica

La computación cuántica es otra de las aplicaciones que requieren de la manipulación del entrelazamiento para su concreción. Esta nueva disciplina comenzó a desarrollarse a partir de los trabajos de Richard Feynman en 1982. Feynman abrió la puerta para reformular un nuevo paradigma computacional motivado directamente por las leyes de la física. En su trabajo pionero, Feynman notó que la simulación computacional de los sistemas cuánticos es altamente ineficiente si se la implementa en cualquier computadora ordinaria. En este contexto, la noción de eficiencia está definida a partir del estudio de la dependencia de los recursos necesarios para resolver un cierto problema con el tamaño de dicho problema. En el caso de la física cuántica es bien sabido que para simular un objeto se necesitan utilizar recursos cuya dependencia con el tamaño del objeto es exponencial. Esto quiere decir que, por ejemplo, para simular el comportamiento de un sistema de 41 spines necesitamos el doble de memoria que para un sistema de 40 spines. De hecho, las computadoras más poderosas de la actualidad no son suficientes para resolver las ecuaciones de la mecánica cuántica para un sistema de alrededor de 50 spines. El motivo de la dificultad en la simulación de los sistemas cuánticos es, precisamente, el entrelazamiento. En cambio, si un objeto compuesto evoluciona visitando estados que nunca están entrelazados, puede ser estudiado eficientemente en una computadora ordinaria. A partir de los trabajos de Feynman un grupo cada vez más grande de científicos comenzaron a trabajar sobre la idea de construir otro tipo de computadoras, en las cuales el "hardware" evolucione de acuerdo a las leyes de la mecánica cuántica. El estudio del poder computacional de este tipo de computadoras, que son denominadas "computadoras cuánticas", es un campo de estudio abierto. Por cierto, no son demasiados los resultados demostrados de manera rigurosa en este terreno. Sin embargo, se sabe que existen ciertos problemas matemáticos que tendrían una solución eficiente utilizando computadoras cuánticas pero que sin embargo no poseen una solución eficiente conocida que pueda ser ejecutada en las computadoras ordinarias. El ejemplo más importante es el problema de la factorización de números enteros. En ese caso, en 1994 Peter Shor demostró la existencia de un algoritmo cuántico que es capaz de encontrar los factores primos de un número entero en un tiempo que depende polinomialmente del tamaño del número (medido por el número de bits que son necesarios para almacenarlo). Por el contrario, no existe ningún algoritmo clásico que permita resolver el problema de la factorización entera en tiempo polinomial.

Cabe aclarar que si bien el entrelazamiento es el principal sospechoso a la hora de buscar responsables del poder de la computación cuántica, este hecho no ha sido demostrado rigurosamente. Por el contrario, es un tema de debate álgido y hasta el momento no se ha demostrado que dicho recurso sea realmente necesario (aunque la mayoría de los investigadores mantiene esa sospecha). La construcción de una computadora cuántica es uno de los desafíos tecnológicos de las próximas décadas.

## VII. Comentarios y metáforas finales

¿Cuál es la imagen del Universo que nos provee la mecánica cuántica? No responderemos completamente esta pregunta aquí sino que solo resumiremos los ingredientes de esta visión a los que nos hemos referido en este trabajo. La mecánica cuántica postula la existencia de propiedades observables de un objeto que son incompatibles entre sí. Esto es algo novedoso y profundo. Para asimilarlo es necesario cambiar radicalmente nuestra visión de la realidad física. En primer término deberíamos admitir que al hablar de las *propiedades de un objeto* podemos generar cierta confusión. Esta terminología nos induce a pensar en algo que es propio del objeto, que le pertenece solamente a él. Por el contrario, la mecánica cuántica establece que aquello a lo que llamamos propiedades (o que más técnicamente denominamos como una "magnitud física observable") es en realidad un canal mediante el cual el objeto interactúa con el mundo que lo rodea. El legado del principio de complementariedad es que los objetos tienen distintas ventanas con las que se conectan con el resto del Universo y que existen ventanas que no son compatibles entre sí. Aquello que llamamos "posición" o "momento" son en realidad idealizaciones que lo único que expresan son distintos mecanismos de interacción (canales) por los cuales los objetos de la naturaleza pueden afectarse mutuamente. Lo que la mecánica cuántica nos enseña es que hay ciertos mecanismos de interacción que son compatibles entre sí y que, por el contrario, hay otros que no lo son. Cuando un objeto interactúa con el mundo que lo rodea mediante el "canal de posición", no puede hacerlo mediante el "canal de momento" y viceversa. En definitiva, la mecánica cuántica nos enseña que los objetos tienen distintas caras y que no todas ellas pueden ser vistas al mismo tiempo. La esencia del principio de complementariedad es esa y ese es un hecho fundamental.

Otra de las enseñanzas de la física cuántica es que el acto de medición no es un hecho pasivo. Probablemente este sea uno de los aspectos más controvertidos de la mecánica cuántica. En efecto, coloca al observador en un lugar diferente del que tradicionalmente le otorgaba la física. Anteriormente se pensaba que las perturbaciones inherentes a la observación podían ser minimizadas. Se pensaba que era posible concebir al acto de observar como una acción asimilable a la de *revelar algo que está escrito en el objeto estudiado*. La mecánica cuántica derribó ese paradigma y lo reemplazó por otro en el que el acto de observar es siempre una interacción. Muchas veces se presenta este hecho como una ventana por la cual puede colarse el subjetivismo. Pero la física cuántica no dice eso sino que establece que el proceso de medición no puede dejar de objetivarse. No puede dejar de describirse como una interacción física. Pero claro, la forma en la que la física cuántica combina esto con la existencia de propiedades incompatibles no puede dejar de sorprendernos. En efecto, si interactuamos con un objeto mediante un cierto canal, determinamos el valor de una de sus propiedades y creamos un estado en el que los valores de sus caras complementarias están completamente indefinidos. Lo sorprendente y anti-intuitivo es que no es posible concebir a este como un estado de *ignorancia* sobre los valores de las caras complementarias. Por el contrario, debe ser tratado como una superposición de todas ellas. Probablemente la lección cuántica que nos resulte más difícil de digerir siga siendo aquella que sintetiza la frase de Asher Peres: *los experimentos que no se realizan no tienen resultados*.

Por último, las predicciones cuánticas para los sistemas compuestos son ciertamente sorprendentes pero a la luz de lo dicho anteriormente no deberían parecerlo tanto. La mecánica cuántica nos dice que podemos encontrar un conjunto de propiedades globales de un sistema compuesto que sean complementarias a todas las propiedades de cualquiera de sus partes. Cuando medimos ese conjunto de propiedades colectivas de un sistema compuesto preparamos al objeto en un estado en el que todas las alternativas de sus facetas complementarias están presentes. Ese es un estado entrelazado en el cual los valores de las propiedades de las partes, que son complementarias con las propiedades medidas, están completamente indefinidos. Es importante destacar que para que este estado mantenga sus propiedades más notables (el entrelazamiento) es vital que permanezca aislado de todo tipo de interacciones con el medio (que típicamente tienen lugar a través de canales locales). Si el objeto permanece aislado y no es afectado por ningún mecanismo que induzca su *decoherencia* entonces seguirá comportándose como un todo. Será un objeto extendido, una unidad no-local, pese a que sus partes se hayan desplazado a lugares distantes. Las manifestaciones del comportamiento cuántico de objetos compuestos cuyas partes entrelazadas están separados por distancias macroscópicas son realmente sorprendentes. El siglo XXI será, sin duda, el siglo donde el estudio, la ingeniería y el aprovechamiento de este tipo de estados darán lugar al desarrollo de novedosas tecnologías cuánticas que, tal vez, contribuyan a que alguna vez la afirmación de Richard Feynman "*nadie entiende la mecánica cuántica*" deje de ser cierta.